\documentclass[11pt,a4paper]{article}
\usepackage{amsmath,amssymb}
\usepackage{epsfig,graphicx}
\usepackage{subfigure}
\usepackage{graphicx}
\usepackage{rotating}
\usepackage{bm}
\usepackage{color}
\usepackage[sort&compress,square]{natbib}

\newcommand{\mg}{m_{\gamma^{\prime}}}
\newcommand{\Bmag}{{\mathbf{B}}}
\newcommand{\Bmaghid}{{\mathbf{B_{hid}}}}

\begin{document}
\date{\mbox{ }}
\title{{\normalsize  IPPP/08/39; DCPT/08/78; DESY-08-069\hfill\mbox{}\hfill\mbox{}}\\
\vspace{2.5cm} \Large{\textbf{Searching Hidden-sector Photons inside a Superconducting Box}}}
\author{Joerg Jaeckel$^{1}$\footnote{{\bf e-mail}: joerg.jaeckel@durham.ac.uk} and Javier Redondo$^{2}$\footnote{{\bf e-mail}: javier.redondo@desy.de}
\\[2ex]
\small{\em $^1$Institute for Particle Physics Phenomenology, Durham University, Durham DH1 3LE, United Kingdom}\\[1.5ex]
\small{\em $^2$Deutsches Elektronen Synchrotron, Notkestra\ss e 85, 22607 Hamburg, Germany}\\[1.5ex]}
\date{}
\maketitle

\begin{abstract}
\noindent
We propose an experiment to search for extra ``hidden-sector" U(1) gauge bosons with gauge kinetic mixing with the ordinary photon, predicted by many extensions of the Standard Model.
The setup consists of a highly sensitive magnetometer inside a superconducting shielding. This is then placed inside a strong (but sub-critical) magnetic field.
In ordinary electrodynamics the magnetic field cannot permeate the superconductor and no field should register on the magnetometer. However, photon -- hidden-sector photon -- photon oscillations would allow to penetrate the superconductor and the magnetic field would ``leak'' into the shielded volume and register on the magnetometer.
Although this setup resembles a classic ``light shining though a wall experiment'' there are two crucial differences. First, the fields are (nearly) static and the photons involved are virtual. Second, the magnetometer directly measures the field-strength and not a probability. This improves the dependence of the signal on the kinetic mixing $\chi\ll 1$ to $\chi^2$ instead of $\chi^4$.
In the mass range $2\,\mu {\rm eV}\lesssim m_{\gamma^{\prime}}\lesssim 200\, {\rm meV}$ the projected sensitivity is in the $\chi\sim 5\times 10^{-9}$ to $\chi\sim 10^{-6}$ range.
This surpasses current astrophysical and laboratory limits by several orders of magnitude -- ample room to discover new physics.
\end{abstract}

In the near future the LHC will commence searching for new particles with masses of the order of a TeV.
This will test many proposed extensions of the Standard Model as, e.g., supersymmetry, large extra dimensions, technicolor to name only a few.
However, many extensions of the Standard Model contain additional hidden sectors that interact only very weakly with ordinary matter. Due to their feeble interactions
even light particles in such hidden sectors may be missed in such a collider experiment. Yet, it may be exactly these hidden sectors that carry crucial information
on how the Standard Model is embedded in a more fundamental theory. This creates the need for complementary experiments.

\begin{figure}
\begin{center}
\includegraphics[width=8.6 cm]{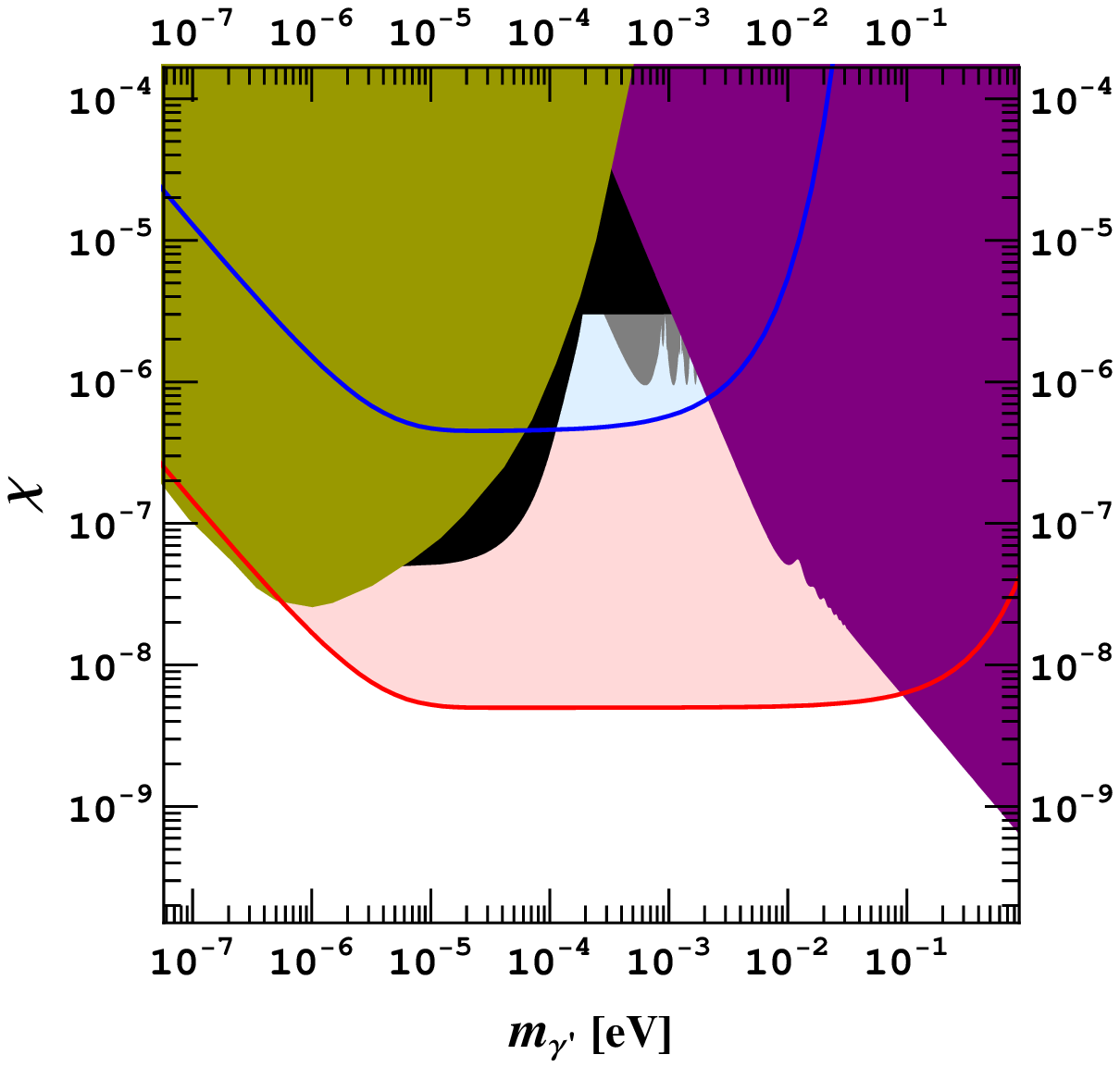}
\end{center}
\vspace{-5ex}
\caption[...]{\small
Current bounds on hidden-sector photons from Coulomb law tests \cite{Williams:1971ms,Bartlett:1988yy} (yellow),
searches of solar hidden photons with the CAST experiment (purple) \cite{Popov:1991,Popov:1999,Andriamonje:2007ew,Redondo:2008aa}
and light-shining-through-walls (LSW) experiments \cite{Cameron:1993mr,Robilliard:2007bq,Chou:2007zz,Ahlers:2007rd,Ahlers:2007qf} (grey)
as well as CMB measurements of the effective number of neutrinos $\Delta N^{\rm eff}_{\nu}$
and the blackbody
nature of the spectrum (black) \cite{Mangano:2006ur,Ichikawa:2006vm,Komatsu:2008hk,Jaeckel:2008fi}.
Improvements of the solar bounds can be achieved using the SuperKamiokande detector or upgrading the CAST experiment \cite{Gninenko:2008pz}.
The region $m_{\gamma^{\prime}}\lesssim 1\, {\rm meV}$ could be tested by an
experiment using microwave cavities \cite{Jaeckel:2007ch}.
The lightly shaded areas bounded by lines give the projected sensitivity for the experiment proposed in this note. The blue area
corresponds to a relatively conservative estimate for the magnetometer sensitivity $\Bmag_{\rm{detect}}\sim 10^{-14}\,{\rm T}$,
and a thickness of the shielding $d\sim 0.1\,{\rm mm}$ -- much greater
than  the theoretical minimum required to have sufficient \mbox{shielding~--,} an external field of $\Bmag_{0}=0.05\, {\rm T}$ is assumed.
The red area is an optimistic scenario, $\Bmag_{\rm{detect}}\sim 5\,10^{-18}\,{\rm T}$, $d\sim 50\,\lambda_{\rm Lon}\sim 1\mu{\rm m}$ and $\Bmag_{0}=0.2\,{\rm T}$.
(For both scenarios we used $L_{1}=10\,{\rm cm}$ for the distance from the magnetic field source to the shield and $L_{2}=5\,{\rm cm}$ for the distance
from the shield to the magnetometer.)
\label{exclu}}
\end{figure}

One 
interesting class of hidden-sector particles is extra U(1) gauge bosons. Indeed, many string theory models
contain hidden-sector U(1)s under which Standard Model particles are uncharged.
At low energies,  the only
renormalizable interaction with the visible sector
can occur via mixing~\cite{Okun:1982xi,Holdom:1985ag,Foot:1991kb} of
the photon $\gamma$ with the hidden sector photon $\gamma^\prime$.
Current experimental bounds are shown in Fig. \ref{exclu}.
Interestingly there is a dip in the sensitivity in the mass region around $1\, {\rm meV}$.
Hidden-sector photons in this mass region could, e.g., explain \cite{Jaeckel:2008fi} the slight excess in the number of relativistic species
observed in combinations of CMB with Lyman-$\alpha$ data \cite{Seljak:2006bg,Cirelli:2006kt} (although this could also be due to systematics \cite{Hamann:2007pi}).
Moreover, one might (wildly) speculate about
possible connections to neutrino masses and dark energy which incidentally also have energy scales around meV.
The experiment proposed in this note is designed to search for hidden-sector photons exactly in this region.

\begin{figure}
\begin{center}
\includegraphics[width=8.6 cm]{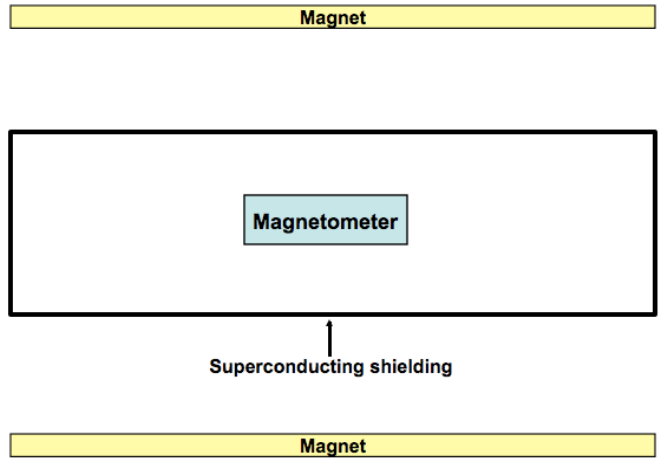}
\end{center}
\vspace{-2ex}
\caption[...]{\small
Sketched setup for the experiment proposed in this note. Ordinary magnetic fields are shielded by the superconductor. However,
if a hidden U(1) field mixes with the ordinary electromagnetic field some of the magnetic field can convert into a hidden magnetic field,
pass through the superconductor and reconvert into an ordinary magnetic field inside the shielding. This field can then be measured by a highly sensitive magnetometer.
\label{magnet}}
\end{figure}

The basic idea of the proposed experiment is very similar to a classic ``light shining through a wall experiment'' .
However, instead of light it uses a static magnetic field and the wall is replaced by superconducting shielding (cf. Fig. \ref{magnet}).
Outside the shielding we have a strong magnetic field. Upon entering the superconductor the ordinary electromagnetic field
is exponentially damped with a length scale given by the London penetration depth $\lambda_{\rm Lon}$. Yet, due to the photon -- hidden photon mixing
a small part of the magnetic field is converted into a hidden magnetic field.
After the superconducting shield is crossed the mixing turns a small fraction of the hidden magnetic field back into an ordinary magnetic field that can be detected by a magnetometer.
Since the magnetometer measures directly the field (and not some probability or power output) the signal is proportional
to the transition amplitude and therefore to the mixing squared, $\chi^2$, instead of being proportional to $\chi^4$.

Let us start from the Lagrangian describing the photon -- hidden-sector photon system at low energies,
\begin{equation}
\label{lagrangian}
{\mathcal{L}}= -\frac{1}{4} F^{\mu\nu}F_{\mu\nu}-\frac{1}{4}B^{\mu\nu}B_{\mu\nu}
-\frac{1}{2}\chi\,F^{\mu\nu}B_{\mu\nu}  +\frac{1}{2}\mg^2 B_\mu B^\mu+j_\mu A^\mu+\frac{1}{2}M_{\rm Lon}^2 A_\mu A^\mu,
\end{equation}
where $F_{\mu\nu}$ is the field strength tensor for the ordinary
electromagnetic {U(1)$_{_\mathrm{QED}}$} gauge field $A^{\mu}$,
$j^\mu$ is its associated current (generated by
electrons, etc.) and $B^{\mu\nu}$ is the
field strength for the hidden-sector {U(1)$_\mathrm{h}$} field
$B^{\mu}$. The first two terms are the standard kinetic terms for
the photon and hidden photon fields, respectively. Because the field
strength itself is gauge invariant for U(1) gauge fields, the third
term is also allowed by gauge and Lorentz symmetry.  This term
corresponds to a non-diagonal kinetic term, a so-called kinetic
mixing~\cite{Holdom:1985ag}. This term is a renormalizable dimension
four term and does not suffer from mass suppressions. It is
therefore a sensitive probe for physics at very high energies.
Kinetic mixing arises in field theoretic \cite{Holdom:1985ag} as
well as in string theoretic setups
\cite{Dienes:1996zr,Lust:2003ky,Abel:2003ue,Batell:2005wa,Blumenhagen:2006ux,Abel:2006qt,Abel:2008ai}.
Typical predicted values for $\chi$ in realistic string
compactifications range between $10^{-16}$ and $10^{-2}$. The second
to last term is a mass term for the hidden photon. This could either
arise from a Higgs mechanism or it could be a St\"uckelberg mass
term~\cite{Stueckelberg:1938}. Finally, the last term corresponds to
the London mass $M_{\rm Lon}=1/\lambda_{\rm Lon}$ the photon
acquires inside a superconductor. In vacuum $M_{\rm Lon}=0$.

From this Lagrangian we obtain the equations of motion,
\begin{eqnarray}
\label{eqspot}
&&\!\!\!\!\!\!\square (A^{\mu}+\chi B^{\mu})+M^{2}_{\rm Lon}A^{\mu}+j^\mu=0
\\\nonumber
&&\!\!\!\!\!\!\square (\chi A^{\mu}+B^{\mu})+m^{2}_{\gamma^{\prime}} B^{\mu}=0.
\end{eqnarray}
Here, we have chosen Lorentz gauge for the massless field (if $M^{2}_{\rm Lon}=0$) and used that the Lorentz condition is automatically enforced for massive fields.

Due to the kinetic mixing, a source $j^{\mu}$ originating from purely standard model fields, i.e. typically electrons, will automatically generate also a small $B^{\mu}$ field:
Very close to a localized source the the field equations \eqref{eqspot} are dominated by the derivative terms and the mass terms can be neglected.
Accordingly the second equation in \eqref{eqspot} can only be fulfilled  if the generated field is proportional to $(A,B)\sim(1,-\chi)$.
Having said this, we will drop the source from now on.

Taking appropriate derivatives in \eqref{eqspot} we get the equations for the field strengths,
\begin{eqnarray}
\label{fieldequations}
&&\!\!\!\!\!\!\square (F^{\mu\nu}+\chi B^{\mu\nu})+M^{2}_{\rm Lon} F^{\mu\nu}=0
\\\nonumber
&&\!\!\!\!\!\!\square (\chi F^{\mu\nu}+B^{\mu\nu})+m^{2}_{\gamma^{\prime}} B^{\mu\nu}=0.
\end{eqnarray}

The essential features can be studied by using an one-dimensional setup: The superconductor fills the 2-3-plane and has a thickness $d$.
The distances from the source of the magnetic field to the superconducting plane and from there to the detector will be called $L_1$ and $L_{2}$, respectively.
Note that the different space-time components of Eq. \eqref{fieldequations} decouple and fulfill the same equations so essentially we can drop the spacetime indices. As an example let us choose a static magnetic field in the 3 direction, i.e. only the 12 components are non-vanishing, $F^{12}=\Bmag$ and $B^{12}=\Bmaghid$. We find,
\begin{eqnarray}
\label{simpleequations}
&&\!\!\!\!\!\!\partial^{2}_{1} (\Bmag+\chi \Bmaghid)-M^{2}_{\rm Lon} \Bmag=0,
\\\nonumber
&&\!\!\!\!\!\!\partial^{2}_{1} (\chi \Bmag+\Bmaghid)-m^{2}_{\gamma^{\prime}} \Bmaghid=0.
\end{eqnarray}
Outside the superconductor $M^{2}_{\rm Lon}=0$ and to lowest non-trivial order in $\chi$ the non-growing solutions are,
\begin{eqnarray}
\label{vacsol}
V_{1}(x)=\left(
           \begin{array}{c}
             1 \\
             0 \\
           \end{array}
         \right)\quad{\rm and}\quad
V_{2}(x)=\left(
        \begin{array}{c}
          -\chi \\
          1 \\
        \end{array}
      \right)\exp(-\mg x)\quad {\rm for}\quad M_{\rm Lon}=0,
\end{eqnarray}
in the $(\Bmag,\Bmaghid)$ basis.
As discussed above, close to the source the magnetic field will be $\sim(1,-\chi)$.
However, we can now see that the small $\Bmaghid$ component is exponentially damped.
Therefore, after a distance $\gg 1/\mg$, the field is $\propto V_1$ and will stay that way until it reaches the superconductor.

Inside the superconductor the solutions are,
\begin{eqnarray}
V^{sc}_{1}(x)=\left(\begin{array}{c}
             1-\frac{m^{2}_{\gamma^{\prime}}}{M^{2}_{\rm Lon}} \\
            -\chi
           \end{array}\right)\exp(-M_{\rm Lon} x)
\quad {\rm and}\quad
V^{sc}_{2}(x)=\left(
                \begin{array}{c}
                 \frac{m^{2}_{\gamma^{\prime}}}{M^{2}_{\rm Lon}-m^{2}_{\gamma^{\prime}}}  \chi\\
                  1 \\
                \end{array}
              \right)\exp(-\mg x).
\end{eqnarray}
Note that $V_2^{sc}$ is damped with $m_{\gamma^{\prime}}$ and not $M_{\rm Lon}$.
For $\mg\ll M_{\rm Lon}$ the mainly hidden photon like $V^{sc}_{2}$ can therefore ``leak'' through the superconducting shielding.

Therefore, for $L_1 \gg 1/\mg $ (distant source) and using $m^{2}_{\gamma^{\prime}}\ll M^{2}_{\rm Lon}$ for simplicity, we have
\begin{eqnarray}
\left(
  \begin{array}{c}
    \Bmag(L_{1}) \\
    \Bmaghid(L_{1}) \\
  \end{array}
\right)=\Bmag_{0}\left(
                    \begin{array}{c}
                      1 \\
                      0 \\
                    \end{array}
                  \right)
=\Bmag_{0}(V^{sc}_{1}(0)+\chi V^{sc}_{2}(0)).
\end{eqnarray}
For $m^{2}_{\gamma^{\prime}}\ll M^{2}_{\rm Lon}$ and
$1/\mg \gg d\gtrsim {\rm few}\times 10 \lambda_{\rm Lon}$ only the component $V^{sc}_{2}(x)$ effectively penetrates the shielding and
\begin{eqnarray}
\left(
  \begin{array}{c}
    \Bmag(L_{1}+d) \\
    \Bmaghid(L_{1}+d) \\
  \end{array}
\right)
\simeq \Bmag_{0} \chi V^{sc}_{2}(d)\simeq\Bmag_{0}\chi\left(         \begin{array}{c}              0 \\                  1 \\  \end{array}  \right) =
\chi^2 V_{1}(0)+\chi V_{2}(0)
.
\end{eqnarray}
At a distance 
$L_2\gg 1/\mg$ after the shielding only the $V_{1}$ component survives and the measurable ordinary magnetic
field approaches $\Bmag_{\rm meas}\equiv\Bmag(L_{1}+d+L_{2})\simeq\Bmag_{0}\chi^2$.

\begin{figure}
\begin{center}
\includegraphics[width=8.6 cm]{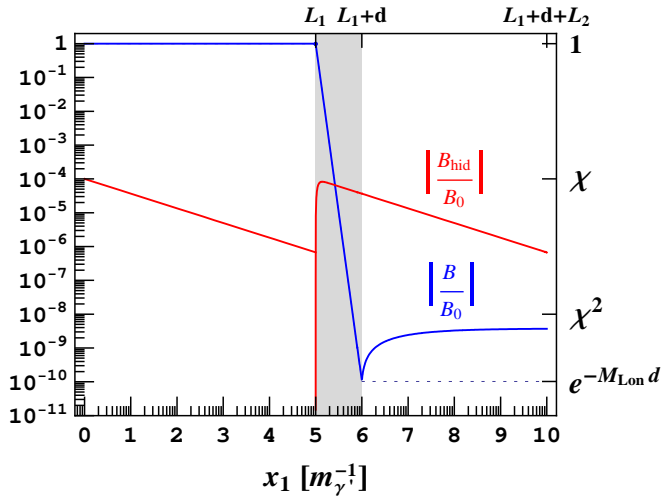}
\end{center}
\vspace{-2ex}
\caption[...]{\small Dependence of the magnetic field $\Bmag$ (blue) and hidden magnetic field $\Bmaghid$ (red)
as a function of the distance from the source ($x_1$) in our 1-dimensional set up.
Both quantities are normalized to the value $\Bmag_0$. The dotted line represents the evolution of $\Bmag$ in absence of the hidden field ($\chi\rightarrow 0$).
Near the external (left) surface of the superconducting shield $\Bmaghid$ changes sign, from negative to positive.
For better visibility we have chosen not too extreme values
$\chi=10^{-4}$, $\mg L_1=\mg L_2=5$, $\mg d=1$ and $23\, \mg = M_{\rm Lon}$.
For $\mg d \ll 1\ll L_{1,2}\mg$ the evolution approaches $\chi^2$, but for the chosen parameters the result is a bit smaller
because the hidden field is a bit damped inside the shielding due to $\mg d=1$.
\label{evo}}
\end{figure}

Keeping somewhat more carefully track of the exponentials we find,
\begin{eqnarray}
\label{final}
\frac{\Bmag_{\rm meas}}{\Bmag_0} \!\!&=&\!\! e^{-M_{\rm Lon} d}+\chi^2 e^{-m_{\gamma^{\prime}} d}
\left[1-e^{-\mg L_1}\right]
\left[1-e^{-\mg L_2}\right]\ . 
\end{eqnarray}
The first part is the standard ``leaking'' expected from the finite thickness of the superconducting shielding, and constitutes the standard model background.
The second part $\sim \chi^2$ is the signal caused by the hidden photons.
The first exponential arises from the damping of the hidden photon field inside the superconducting shield whereas the terms in brackets account for the fact that the conversion of photons into hidden
photons and vice versa takes place over length scales $\sim 1/m_{\gamma^{\prime}}$.
The evolution of $\Bmag$ and $\Bmaghid$ from the source to the detector is sketched in Fig. \ref{evo}.

Let us now turn to the sensitivity of such an experiment. Magnetic field strengths $\Bmag_{0}$ of the order $(1-5)\,{\rm T}$ can be reached in the laboratory.
However, we have to stay below the critical field strength of the superconductor. In most materials this ranges between $0.01\,{\rm T}$ and $0.2\,{\rm T}$~\cite{Rohlf} although fields as high as $1$ T can be shielded in certain cases (cf., e.g. \cite{Cavallin}).
Modern magnetometers \cite{gravityb,robbes,Allred} can can detect magnetic fields as low as $\Bmag_{\rm detect}=5\times 10^{-18}\, {\rm T}$ and $\Bmag_{\rm detect}=1\times 10^{-13}$
seems relatively conservative. Accordingly we can expect a sensitivity in the $\chi\sim 10^{-6}$ to $5\times 10^{-9}$ range\footnote{The shielding
of a ${\cal O}(0.1\,{\rm T})$ magnetic field down to ${\cal O}(10^{-18}\,{\rm T})$ might be an experimental challenge (see, however, \cite{gravityb}).}.
Concerning the use of SQUIDs as magnetometers a comment is in order. SQUIDs can only measure \emph{changes} in the magnetic flux. Two possibilities to solve
this problem are to modulate the strength of the external magnetic field or to change its direction (for example by rotating it). As long as the timescale for the
change in the magnetic field is long compared to $\mg$ the above considerations remain unchanged.
In addition, time-modulated magnetic fields might also be helpful in reducing the background.

Finally, we have to ask which mass scales we can probe in the experiment. From \eqref{final} we can read off that optimal sensitivity requires
$L_{1},L_{2}\gg 1/\mg\gg d\gg 1/M_{\rm Lon}$.
Typical London penetration depths $\lambda_{\rm Lon}=1/M_{\rm Lon}$ are of the order of $(20-100)\,{\rm nm}$ (cf,. e.g., \cite{Kittel}).
To avoid fields leaking directly through the shielding (without having to convert into hidden fields) at the $10^{-20}$ level we need
$d\gtrsim 50\, \lambda_{\rm Lon}\sim (1-5)\,\mu {\rm m}$.
The requirement $1/\mg \gg d$ then allows, in principle, to search for masses up to $(0.2-0.04)\,{\rm eV}$.
With thicker than the minimal required shielding the experiment will be sensitive only to smaller masses.
Finally, the maximal size of the magnetic field and the shielded volume will typically be of the order ${\rm few}\times 10\,{\rm cm}$.
This suggests a lower limit $\sim 2\,\mu{\rm eV}$. Overall we have,
\begin{equation}
{\rm Sensitivity:}\quad \chi \lesssim 5\times 10^{-9}- 10^{-6}\quad {\rm for}\quad 2\,\mu{\rm eV}\lesssim \mg \lesssim (0.2-0.04)\,{\rm eV}.
\end{equation}
Two examples for the expected sensitivity are shown in Fig. \ref{exclu}.

\emph{In conclusion:} We have proposed a simple experiment to search for massive hidden photons.  The experiment could improve
upon existing bounds by several orders of magnitude in a mass range $2\,\mu{\rm eV}\lesssim \mg \lesssim (0.2-0.04)\,{\rm eV}$
where existing bounds are relatively weak. It therefore bears significant discovery potential for hidden sector physics at the meV scale.
In particular, it could test recent speculations \cite{Jaeckel:2008fi} that the excess in the observed number of relativistic degrees of freedom at decoupling of the CMB originates from a resonant production of hidden photons in the early universe.

\section*{Acknowledgments}
We would like to thank Holger Gies and Alvar Sanchez for very helpful and interesting discussions.

\end{document}